\documentclass[conference]{IEEEtran}
\IEEEoverridecommandlockouts
\pdfoutput=1

\usepackage{amsmath,amssymb,amsfonts}
\usepackage{cite}
\usepackage{algorithm}
\usepackage[noend]{algpseudocode}
\usepackage{graphicx}
\usepackage{textcomp}
\usepackage{xcolor}
\usepackage[normalem]{ulem}

\newcommand{\calA}{\ensuremath{\mathcal{A}}}

\newcommand{\calH}{\ensuremath{\mathcal{H}}}

\newcommand{\calM}{\ensuremath{\mathcal{M}}}
\newcommand{\calN}{\ensuremath{\mathcal{N}}}

\newcommand{\calQ}{\ensuremath{\mathcal{Q}}}

\newcommand{\calY}{\ensuremath{\mathcal{Y}}}

\DeclareMathOperator*{\argmax}{arg~max}

\newcommand{\R}{\ensuremath{\mathbb{R}}}
\newcommand{\E}{\ensuremath{\mathbb{E}}}
\newcommand{\Z}{\ensuremath{\mathbb{Z}}}

\newcommand{\te}{\ensuremath{\tilde{e}}}

\newcommand{\tx}{\ensuremath{\tilde{x}}}

\newcommand{\tP}{\ensuremath{\tilde{P}}}

\newcommand{\tQ}{\ensuremath{\tilde{Q}}}

\newcommand{\tL}{\ensuremath{\tilde{L}}}

\newcommand{\hx}{\ensuremath{\hat{x}}}
\newcommand{\he}{\ensuremath{\hat{e}}}

\newcommand{\hL}{\ensuremath{\hat{L}}}

\newcommand{\hP}{\ensuremath{\hat{P}}}

\newcommand{\bU}{\ensuremath{\bar{U}}}

\newcommand{\bone}{\ensuremath{\mathbf{1}}}

\newcommand{\dg}{\ensuremath{\dot{g}}}

\newcommand{\dx}{\ensuremath{\dot{x}}}

\newcommand{\ddg}{\ensuremath{\ddot{g}}}

\newcommand{\tr}{\ensuremath{\textmd{tr}}}

\newtheorem{remark}{Remark}

\newtheorem{definition}{Definition}
\newtheorem{proposition}{Proposition}
\newtheorem{theorem}{Theorem}
\newtheorem{lemma}{Lemma}

\makeatletter
\def\BState{\State\hskip-\ALG@thistlm}
\makeatother

\def\BibTeX{{\rm B\kern-.05em{\sc i\kern -.025em b}\kern-.08em
    T\kern-.1667em\lower.7ex\hbox{E}\kern-.125emX}}
\begin{document}

\title{\bf \LARGE A Game-Theoretic Approach to Secure Estimation and Control for Cyber-Physical Systems with a Digital Twin}
\author{
\IEEEauthorblockN{1\textsuperscript{st} Zhiheng Xu}
\IEEEauthorblockA{School of Electrical and Electronic Engineering \\
Nanyang Technological University, Singapore \\
Email: zhiheng.xu688@gmail.com}
\and
\IEEEauthorblockN{2\textsuperscript{nd} Arvind Easwaran}
\IEEEauthorblockA{School of Computer Science and Engineering \\
Nanyang Technological University, Singapore \\
Email: arvinde@ntu.edu.sg}

\thanks{This work was supported by Delta-NTU Corporate Lab for Cyber-Physical Systems with funding support from Delta Electronics Inc. and the National Research Foundation (NRF) Singapore under the Corp Lab@University Scheme.}
}

\maketitle

\begin{abstract}
Cyber-Physical Systems (CPSs) play an increasingly significant role in many critical applications. These valuable applications attract various sophisticated attacks. This paper considers a stealthy estimation attack, which aims to modify the state estimation of the CPSs. The intelligent attackers can learn defense strategies and use clandestine attack strategies to avoid detection. To address the issue, we design a Chi-square detector in a Digital Twin (DT), which is an online digital model of the physical system. We use a Signaling Game with Evidence (SGE) to find the optimal attack and defense strategies.  Our analytical results show that the proposed defense strategies can mitigate the impact of the attack on the physical estimation and guarantee the stability of the CPSs. Finally, we use an illustrative application to evaluate the performance of the proposed framework.
\end{abstract}

\begin{IEEEkeywords}
Cyber-Physical Systems, Data-Integrity Attack, Signaling Game with Evidence, Digital Twin.
\end{IEEEkeywords}

\section{Introduction}

Cyber-Physical Systems (CPS) integrate physical components (e.g., sensors, actuators, and controllers), computational resources, and networked communications \cite{kim2012cyber}. The integration with networked communications highly enhances the flexibility and scalability of CPSs in various applications, such as large-scale manufacturing systems \cite{wang2015current}, intelligent transport systems \cite{xiong2015cyber}, and smart grid infrastructure \cite{mo2011cyber}. However, the valuable applications of CPSs attract many sophisticated attacks. A well-known example is the Stuxnet, a computer virus compromises Supervisory control and data acquisition (SCADA) of industrial systems \cite{langner2011stuxnet}. Besides, we have witnessed many other control-system-related attacks, such as Maroochy Water attack \cite{slay2007lessons}, Unmanned Aerial Vehicle's (UAV) GPS spoofing attack \cite{shepard2012drone}, and German Steel Mill cyber attack \cite{lee2014german}.

Due to the increasing number of security-related incidents in CPSs, many researchers have studied the features of these attacks and developed relevant defense strategies. Among many attack models, we focus on data-integrity attacks, where the attackers modify the original data used by the system or inject unauthorized data to the system \cite{mo2012integrity}. The data-integrity attacks can cause catastrophic consequences in CPSs. For instance, the fake data may deviate the system to a dangerous trajectory or make the system oscillate with a significant amplitude, destabilizing the system. Therefore, how to mitigate the impact of data-integrity attacks becomes a critical issue in the security design of CPSs.

One typical data-integrity attack for CPSs is the Sensor-and-Estimation (SE) attack, where the attackers tamper the sensing or estimated information of the CPSs \cite{pa2015control}.  Given the SE attack, Fawzi et al. \cite{fawzi2014secure} have studied a SE attack and proposed algorithms to reconstruct the system state when the attackers corrupt less than half of the sensors or actuators. Miroslav Pajic et al. \cite{pajic2016attack} have extended the attack-resilient state estimation for noisy dynamical systems. Based on Kalman filter, Chang et al. \cite{chang2018secure} have extended the secure estimation of CPSs to a scenario in which the set of attacked nodes can change over time. However, to recover the estimation, the above work requires a certain number of uncorrupted sensors or a sufficiently long time. Those approach might introduce a non-negligible computational overhead, which is not suitable for time-critical CPSs, e.g., real-time systems. Besides, since all the senors do not have any protection, the attacker might easily compromise the a large number of sensors, violating the assumptions of the above work.

Instead of recovering the estimation from SE attacks, researchers and operators also focus on attack detection \cite{ding2018survey}. However, detecting a SE attack could be challenging since the attackers' strategies become increasingly sophisticated. Pasqualetti et al. \cite{pa2013attack} have identified the conditions of undetectable SE attacks for CPSs. Using the conventional statistic detection theory, e.g., Chi-square detection, may fail to discover an estimation attack if the attackers launch a stealthy attack \cite{mo2015performance}. Yuan Chen et al. \cite{chen2017optimal} have developed optimal attack strategies, which can deviate the CPSs subject to detection constraints. Hence, the traditional detection theory may not sufficiently address the stealthy attacks in which the attackers can acquire the defense information. Besides, using classical cryptography to protect CPSs will introduce significant overhead, degrading the control performance of delay-sensitive CPSs \cite{xu2018cross}.

The development of Digital Twin (DT) provides essential resources and environments for detecting sophisticated attacks. A DT could be a virtual prototype of a CPS, reflecting partial or entire physical processes \cite{luo2019digital}. Based on the concept of DT, researchers have developed numerous applications for CPSs \cite{tao2018digital}. The most relevant application to this paper is using a DT to achieve fault detection \cite{bi2016com}. Similarly, we use the DT to monitor the estimation process, mitigating the influence of the SE attack.

\begin{figure}
  \centering
  \includegraphics[width=6.2cm]{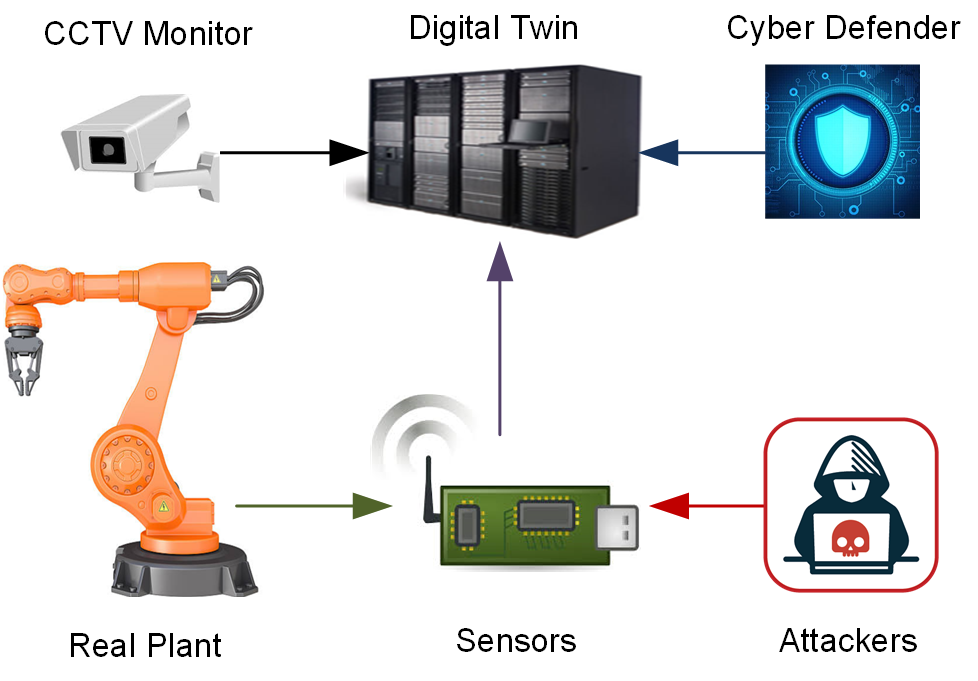}
  \caption{The Stealthy Estimation Attack and the Defense Mechanism based on a Digital Twin: the attacker aims to modify the estimation results in a CPS, while the Digital Twin protects the system by monitoring the results.}\label{fig:exampleDT}
  \vspace{-4mm}
\end{figure}

In this paper, we focus on a stealthy estimation attack, where the attackers know the defense strategies and aim to tamper the estimation results without being detected. Fig. \ref{fig:exampleDT} illustrates the underlying architecture of the proposed framework. To withstand the attack, we design a Chi-square detector, running in a DT. The DT connects to a group of protected sensing devices, collecting relevant evidence. We use cryptography (e.g., digital signature or message authentication code) to preserve the evidence from the attack. Hence, the DT can use the evidence to monitor the estimation of the physical systems. The cryptographic overhead will not affect physical performance since the execution of the real plant does not depend on the DT.

\begin{figure}[thp]
  \centering
  \includegraphics[width=7.2cm]{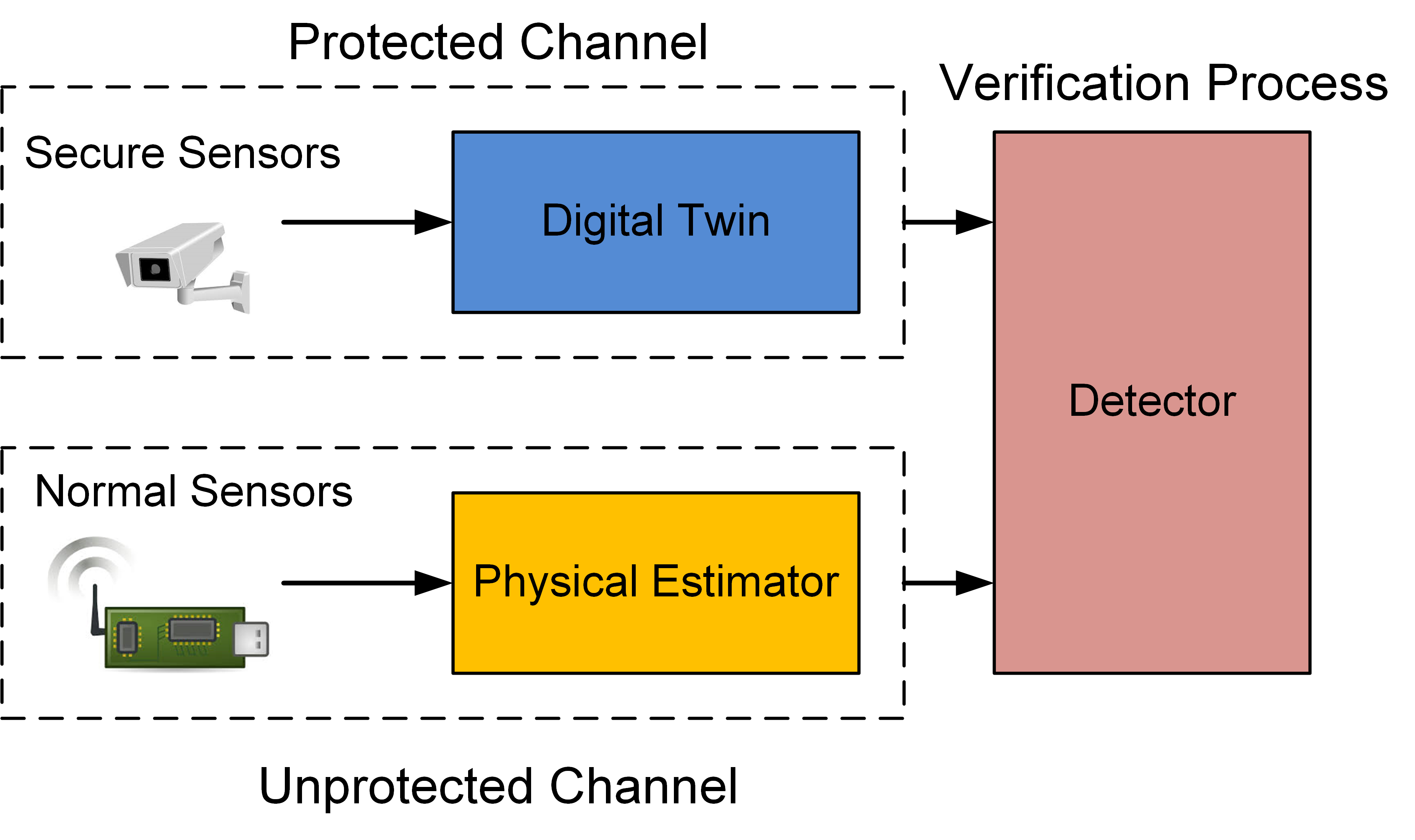}
  \caption{The Architecture of the Framework: the Digital Twin has two channels to obtain the sensing information of the plant: one is secure, and the other is not secure. The secure channel provides less accurate data with a heavy computational overhead. DT only uses the secure channel to run the verification, and the computational overhead has negligible impact on the physical performance since these two channels are independent. }\label{fig:comparsion}
  \vspace{-2mm}
\end{figure}

Different from the work \cite{fawzi2014secure,pajic2016attack,chang2018secure}, we have designed two independent channels, i.e., one is protected by standard cryptography, and the other one is the general sensing channel. Fig. \ref{fig:comparsion} illustrates the structure of the framework. The main advantage of this structure is the cryptographical overhead will affect the control performance of the physical system due to the independency between these two channels.

To analyze whether a stealthy attack can bypass DT's detector, we use game theory to find the optimal attack and defense strategies. Game theory has been an essential tool in designing security algorithms since we can use it to search for the optimal defense strategies against intelligent attackers \cite{manshaei2013game}. One related game model that can capture the detection issue is the Signaling Game (SG) \cite{shen2011signaling}. However, instead of using the SG, we use a Signaling Game with Evidence (SGE), presented in \cite{pawlick2018modeling}, to protect the system from the attack. In an SGE, the DT's detector will provide critical evidence to explore the stealthy attack. After integrating DT's detector with CPSs, we use an SGE to study the stealthy attack and develop optimal defense strategies based on the equilibria of the SGE. Our analytical results show that the stealthy attackers have to sacrifice the impact on the physical system to avoid detection.

We organize the rest of the paper as follows. Section \ref{sec:probF} presents the problem formulation in which we identify the control problem, attack model, and a signaling game framework. Section \ref{sec:main} analyzes the equilibrium of the game framework and the stability of the CPSs. Section \ref{sec:simRes} uses the experimental results to evaluate the performance of the proposed defense mechanism. Finally, Section \ref{sec:con} concludes the entire paper.

\section{System Modelling and Characterization}\label{sec:probF}

In this section, we first introduce the dynamic model of a CPS. Secondly, we define a stealthy estimation attack model. Based on a Digital Twin (DT), we design a Chi-square detector to monitor the estimation process. Finally, we define a Signaling Game with Evidence (SGE) to characterize the features of the stealthy attack.

\subsection{System Model and Control Problem of a CPS}

Suppose that the physical layer of a CPS is a control system. We assume that the control system can be linearized as a linear discrete-time system, given by
\begin{eqnarray}
  x_{k+1} &=& Ax_k + Bu_k + w_k, \label{eq:sysModel}  \\
  y_k &=& C x_k + v_k,           \label{eq:senModel}
\end{eqnarray}
where $k\in\Z_{+}$ is the discrete-time instant; $x_k\in\R^{n_x}$ is the system state with an initial condition $x_0\thicksim\calN(0_{n_x}, \Sigma_x)$, and $\Sigma_x\in\R^{n_x\times n_x}$ is the covariance matrix; $u_k\in\R^{n_u}$ is the control input; $y_k\in\R^{n_y}$ is the sensor output; $w_k\in\R^{n_x}$ and $v_k\in\R^{n_y}$ are additive zero-mean Gaussian noises with covariance matrices $\Sigma_w$ and $\Sigma_v$ with proper dimensions; $A$, $B$, and $C$ are constant matrices with proper dimensions.

Given system model (\ref{eq:sysModel}), we design a control policy $\mu:\R^{n_x}\rightarrow\R^{n_u}$ by minimizing the following expected linear quadratic cost function, i.e.,
\begin{eqnarray}
  J_{\text{LQG}} = \lim_{N\rightarrow \infty}\sup \E\biggl\{ \frac{1}{N}
  \sum_{k=0}^{N-1}\biggl( x^T_kQx_k + u^T_kRu_k\biggl)\biggl\}, \label{eq:costLQG}
\end{eqnarray}
where $Q\in\R^{n_x\times n_x}$ and $R\in\R^{n_u\times n_u}$ are positive-definite matrices.

Note that the controller cannot use state $x_k$ directly, i.e., we need to design an observer to estimate $x_k$. Hence, minimizing function (\ref{eq:costLQG}) is a Linear Regulator Gaussian (LQG) problem. According to the separation principle, we can design the controller and state estimator separately. The optimal control policy $\mu:\R^{n_x}\rightarrow \R^{n_u}$ is given by
\begin{eqnarray}
  \mu_k(x_k) &:= K_kx_k, \text{~with~} K_k:= - (R+B^TV_{k}B)B^TV_{k}, \label{eq:conLaw}
\end{eqnarray}
where $V_k\in\R^{n_x\times n_x}$ is the solution to the linear discrete-time algebraic Riccati equation
\begin{eqnarray}
  V_{k+1} = Q + A^TV_{k} (A-BK_k), \text{~with~} V_0 = I_{n_x},
\end{eqnarray}
and $I_{n_x}\in\R^{n_x\times n_x}$ is an identity matrix.

We assume that $(A,B)$ is stabilizable. Then, $V_k$ will converge to a constant matrix $V$ when $k$ goes to infinity, i.e.,
\begin{eqnarray}
  \lim_{k\rightarrow\infty} V_k = V, \text{~and~}
  \lim_{k\rightarrow\infty} \mu_k(x) = \mu(x) = Kx. \nonumber
\end{eqnarray}

In the next subsection, we will use a Kalman filter to estimate $x_k$ such that controller (\ref{eq:conLaw}) uses this estimated value to control the physical system.

\subsection{Kalman Filter Problem}

To use controller (\ref{eq:conLaw}), we need to design an estimator. Let $\hx_k\in\R^{n_x}$ be the estimation of $x_k$ and $\he_k:=\hx_k - x_k$ be the error of the estimation. Given the observation $\calY_{k-1}:=\{y_0, y_1, \dots, y_{k-1}\}$, we aim to solve the following Kalman filtering problem, i.e.,
\begin{eqnarray}
  \min_{\hx_k\in\R^{n_x\times n_x}} \ \E[(\hx_k - x_k)^T(\hx_k - x_k)|\calY_{k-1}]. \label{eq:estProb}
\end{eqnarray}

To solve (\ref{eq:estProb}), we need the following lemma, which characterizes a conditioned Gaussian distribution.
\begin{lemma} \cite{papoulis2002probability} \label{lem:conGau}
  If $a\in\R^{n_a}$, $b\in\R^{n_b}$ are jointly Gaussian with means $\bar{a}$, $\bar{b}$ and covariances $\Sigma_{a}$, $\Sigma_{b}$, and $\Sigma_{ab} = \Sigma_{ba}^T$, then given $b$, distribution $a$ is a Gaussian with
  \begin{eqnarray}
    \E[a|b] &=& \bar{a} + \Sigma_{ab}\Sigma^{-1}_{b}(b - \bar{b}), \nonumber \\
    \text{Cov}[a|b] &=& \Sigma_{a}
    - \Sigma_{ab}\Sigma^{-1}_{b}\Sigma_{ba}. \nonumber
  \end{eqnarray}
\end{lemma}

To use Lemma \ref{lem:conGau}, we define the covariance matrix of $\he_k$ as
\begin{eqnarray}
   \hP_k &:=& \E[\he_k \he^T_k|y_{k-1}]
   = \E[(\hx_k - x_k) (\hx_k - x_k)^T|y_{k-1}], \nonumber
\end{eqnarray}
with $\hP_0 = \Sigma_x$. Using the results of Lemma \ref{lem:conGau}, we compute the optimal estimation iteratively, i.e.,
\begin{eqnarray}
  \hx_k = A_K\hx_{k-1} + \hL_k(y_{k-1} - C\hx_{k-1}),
  \label{eq:Kalman}
\end{eqnarray}
where $A_K := (A+BK)$. Gain matrix $\hL_k\in\R^{n_x\times n_y}$ and covariance $\hP_k\in\R^{n_x\times n_x}$ are updated using
\begin{eqnarray}
  \hL_k &:=& A_K\hP_kC^T(\Sigma_v + C\hP_kC^T)^{-1}, \nonumber \\
  \hP_{k+1} &:=& \Sigma_w + (A_K - \hL_kC)\hP_kA^T_K. \nonumber
\end{eqnarray}
Assuming that $(A, C)$ is detectable, we can obtain that
\begin{eqnarray}
  \lim_{k\rightarrow \infty} \hP_k = \hP^*, \label{eq:KalCov}
\end{eqnarray}
where $\hP^*\in\R^{n_x\times n_x}$ is a constant positive-definite matrix.

\subsection{Stealthy Estimation Attack}

CPSs face an increasing threat in recent years. Numerous attack models for CPSs or networked control systems (NCSs) have been introduced in \cite{teixeira2012attack}. Among those attacks, one major attack is the data-integrity attack, where the attacker can modify or forge data used in the CPSs. For example, Liu et al. \cite{liu2011false} have studied a false-data-injection attack that can tamper the state estimation in electrical power grids.
\begin{figure}[thp]
  \centering
  \includegraphics[width=6.6cm]{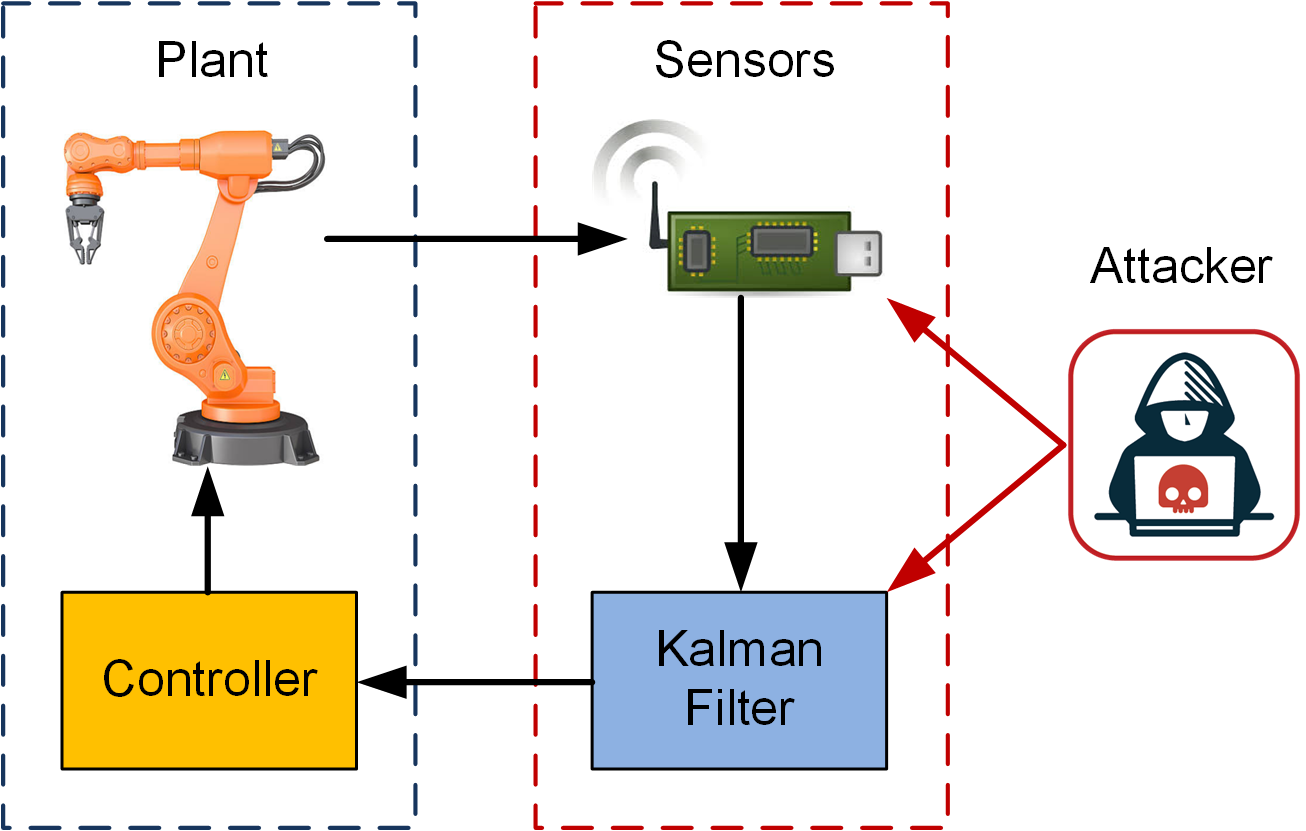}
  \caption{The Stealthy Estimation Attack: intelligent attacker aims to deviate the state by modifying the estimation results.  }\label{fig:attackModel}\vspace{-2mm}
\end{figure}

To mitigate the impact of a data-integrity attack on the state estimation, researchers have designed a Chi-square detector to monitor the estimation process. However, Yilin Mo et al. \cite{mo2015performance} have analyzed stealthy integrity attack, which can pass the Chi-square detector by manipulating the perturbation of the injected data. Therefore, the main challenge is that the conventional fault detectors fail to protect the system from a stealthy attack. One real attack that can achieve the objective is the Advanced Persistent Threat (APT) \cite{tankard2011advanced}, which can compromise a cyber system by executing a zero-day exploration to discover the vulnerabilities.

In our work, we consider an intelligent attacker who can launch a stealthy estimation attack to tamper the estimation results. Fig. \ref{fig:attackModel} illustrates how the attacker achieves its objective. The attacker can either modify the data in the sensors or the data in the estimator. Besides tampering the estimation results, the attacker is also aware of the intrusion detector. The attacker can know the defense strategy and play a stealthy attack to remaind unknown.

In the next subsection, based on the Digital Twin (DT), we design a cyber defender to withstand the stealthy estimation attack and discuss the benefits introduced by the DT. After presenting the game model, we will discuss the optimal defense strategies explicitly in Section \ref{sec:main}.

\subsection{Digital Twin for the CPS}

As mentioned above, an intelligent attacker can learn the defense strategy and launch a stealthy estimation attack, which can modify the estimation results without being detected by the conventional detector, e.g., a Chi-square detector.
To resolve the issue, we aim to design an advanced detector based on a Digital Twin (DT). After that, we use a game-theoretical approach to develop an optimal defense strategy.

Given the system information, we design a DT with the following dynamics
\begin{eqnarray}
    \tx_{k} &=& A_K\tx_{k-1} + \tL_{k}(z_{k-1} - D\tx_{k-1}), \label{eq:estDT} \\
        z_k &=& D x_k + d_k, \nonumber
\end{eqnarray}
where $\tx_k\in\R^{n_x}$ is the DT's estimation of state $x_k$; $z_k\in\R^{n_z}$ is the DT's observation;  $D\in\R^{n_z\times n_x}$ is a constant matrix, $d_k$ is a Gaussian noise with a covariance matrix $\Sigma_d\in\R^{n_z}$.

Similar to problem (\ref{eq:estProb}), we compute $\tL_k$ using the following iterations:
\begin{eqnarray}
  \tL_k &=& A_K\tP_kD^T(\Sigma_{d} + D\tP_{k}D^T)^{-1}, \nonumber \\
  \tP_{k+1} &=& \Sigma_w + (A_K - \tL_{k} D)\tP_{k}A^T_F, \nonumber
\end{eqnarray}
where $\tP_0 = \Sigma_x$ and $\tP_k$ is defined by
\begin{eqnarray}
  \hP_k &:=&\E[(\tx_k - x_k)(\tx_k - x_k)^T|z_{k-1}]
   = \E[\te_k \te^T_k|z_{k-1}]. \nonumber
\end{eqnarray}
where $\te_k:=\tx_k - x_k$ is the DT's estimation error. We also assume that $(A_K, D)$ is detectable, i.e., we have
\begin{eqnarray}
  \lim_{k\rightarrow \infty} \tP_k = \tP^*. \label{eq:DTcov}
\end{eqnarray}

\begin{figure}
  \centering
  \includegraphics[width=6.8cm]{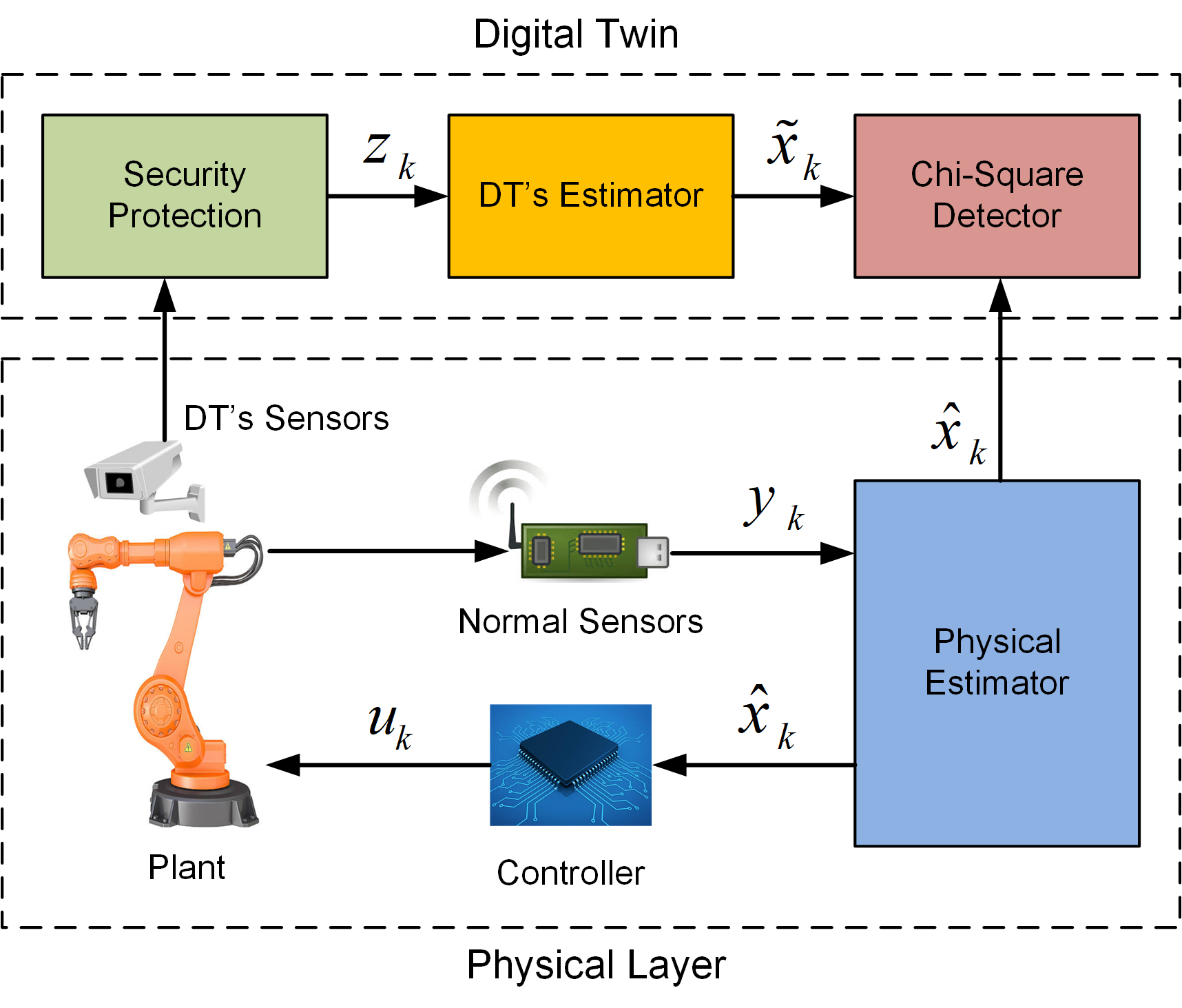}
  \caption{The CPS with a DT: the DT uses a secure observation $z_k$ to obtain an estimation $\tx_k$; given $\tx_k$, we use a Chi-square detector monitor estimation result $\hx_k$.}\label{fig:estDT}
  \vspace{-4mm}
\end{figure}

Fig. \ref{fig:estDT} illustrates the architecture of a CPS with a DT. We summarize the main differences between Kalman filter (\ref{eq:Kalman}) and the DT's estimator (\ref{eq:estDT}) as follows. Firstly, the Kalman filter will use all available sensing information $y_k$ to obtain estimation $\hx_k$. While the DT's estimator just uses a minimum sensing information $z_k\in\R^{n_z}$ to predict $x_k$ as long as $(A, D)$ is detectable, i.e., $n_y \ge n_z$. This feature reduces the dimension of $z_k$, making it easier to protect $z_k$. Secondly, we do not require a high accuracy for $z_k$, since we only use $z_k$ for attack detection. Hence, in general, $\hP^*$ and $\tP^*$ satisfy the condition that $\tr(\hP^*) \le \tr(\tP^*)$, where $\tr(P)$ is the trace of matrix $P$.

Thirdly, we do not use any cryptography to protect $y_k$ since the overhead introduced by the encryption scheme will degrade the performance of the physical system. However, we can use cryptography, such as Message Authentication Code (MAC) \cite{bellare2000security} or Digital Signature (DS) \cite{merkle1989certified}, to protect the integrity of $z_k$. The overhead caused by the cryptography will not affect the physical system, because it does depend on $z_k$. Besides, we can put the DT into a supercomputer or a cloud to resolve the overhead issue.

To sum up, $z_k$ is an observation that is less accurate but more secure than $y_k$. Given the distinct features of $y_k$ and $z_k$, we use $y_k$ to estimate $x_k$ for the physical control and use $z_k$ for the detection in the DT.

Given DT's estimator, we construct a Chi-square detector to monitor estimation result $\hx_k$ at each time $k$. As illustrated in Fig. \ref{fig:estDT}, we build the detector in the DT by comparing $\tx_k$ and $\hx_k$. The Chi-square detector generates a detection result $q_k\in\calQ:=\{0, 1, 2\}$ at time $k$, where $q_k=0$ means the result is qualified, $q_k=1$ means the result is unqualified, $q_k=2$ means the result is detrimental. When $q_k=2$, the DT should always reject the estimation and send an alarm to the operators.

To design the detector, we define $\phi_k:= \tx_k - \hx_k$. Since $\tx_k$ and $\hx_k$ are Gaussian distributions, $\phi_k$ is a also Gaussian distribution with a zero-mean vector and a covariance matrix, i.e., $\phi\thicksim \calN(0_{n_x}, \Sigma_{\phi})$. Furthermore, we define that
\begin{eqnarray}
  \chi^2_k := (\tx_k - \hx_k)^T\Sigma^{-1}_{\phi}(\tx_k - \hx_k). \label{eq:ChiSqu}
\end{eqnarray}
Then, $\chi^2_k$ follows a Chi-square distribution. We define a Chi-square detector as the following:
\begin{eqnarray}
  q_k = f_q(m_k) :=
        \left\{
          \begin{array}{ll}
            0, & \text{~if~} \chi^2_k \le \rho_1; \vspace{1mm}\\
            1, & \text{~if~} \chi^2_k \in(\rho_1, \rho_2]; \vspace{1mm}\\
            2, & \text{~if~} \chi^2_k > \rho_2; \\
          \end{array}
        \right.  \label{eq:CSdetector}
\end{eqnarray}
where $\rho_1, \rho_2$ are two given thresholds, and they satisfy that $\rho_2>\rho_1>0$; $f_q:\R^{n_x}\rightarrow\calQ$ is the detection function.

Using the above Chi-square detector, we can achieve fault detection. However, the work \cite{mo2015performance} has shown that intelligent attackers can constrain the ability of the Chi-square detector by manipulating the amount of injected data. In the following subsection, we will introduce a stealthy sensor attack that aims to remain stealthy while modifying the estimation.

\begin{figure}
  \centering
  \includegraphics[width=6.7cm]{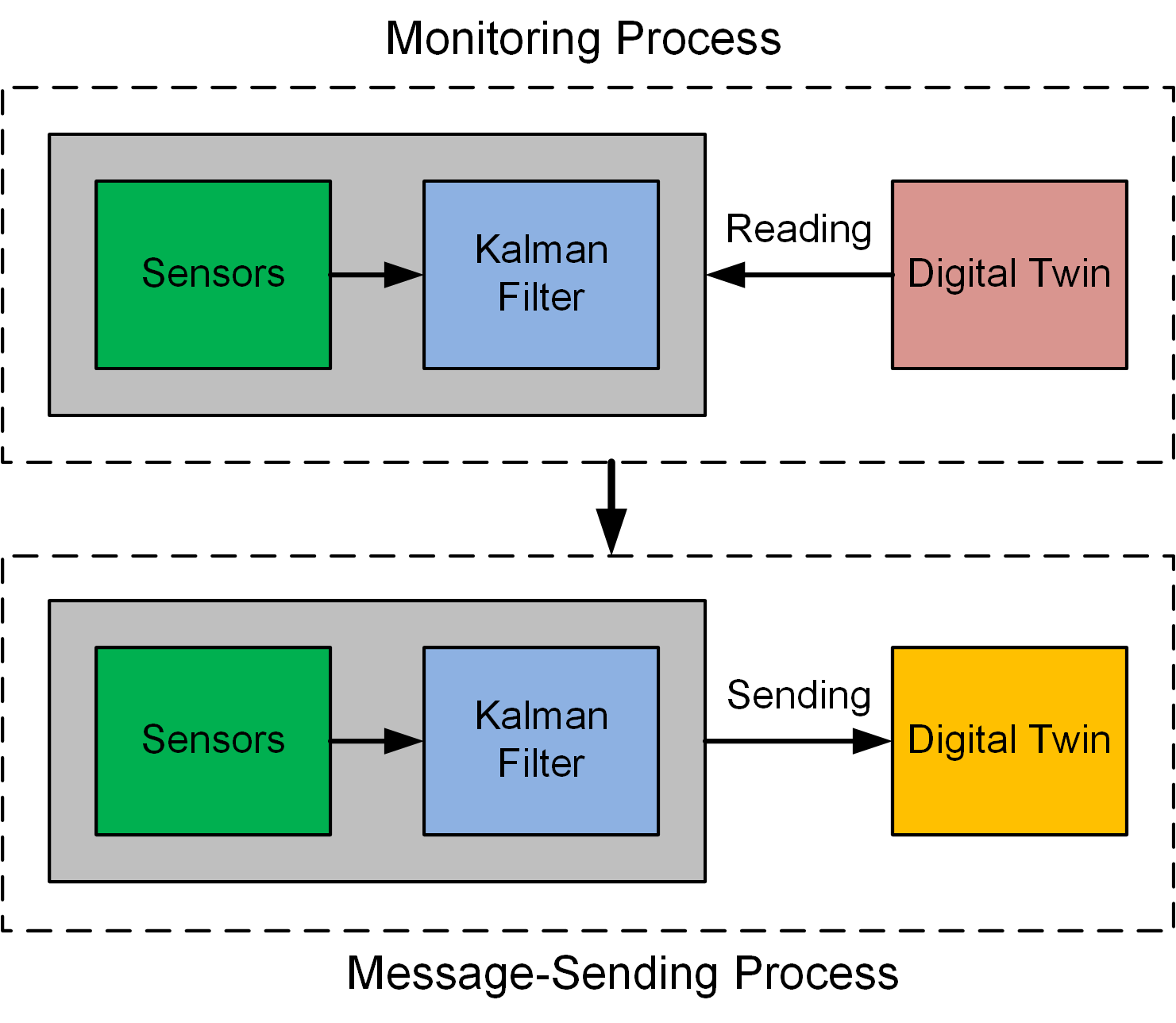}
  \caption{The DT's Monitoring Process: we can view the DT's monitoring process as a message-sending process, i.e., the estimator sends a message to the DT for verification.}\label{fig:megSend}
  \vspace{-4mm}
\end{figure}

\subsection{General Setup of Signaling Game with Evidence}

Due to the existence of attacks, the DT's might not be able to monitor the actual value estimation directly. Instead, the DT's can read a message provided by the estimator. According to our attack model, the attacker can compromise the estimator. Hence, the estimator can have two identities, i.e., a benign estimator or a malicious estimator. The DT aims to verify the estimator's identity by monitoring the estimation results. As shown in Fig. \ref{fig:megSend}, we can view DT's monitoring process as a message-sending process, i.e., the estimator sends an estimation result to the DT for verification. To capture the interactions between the estimator and DT, we will formally define a Signaling Game with Evidence (SGE) as follows.

\begin{figure}[thp]
  \centering
  \includegraphics[width=6.0cm]{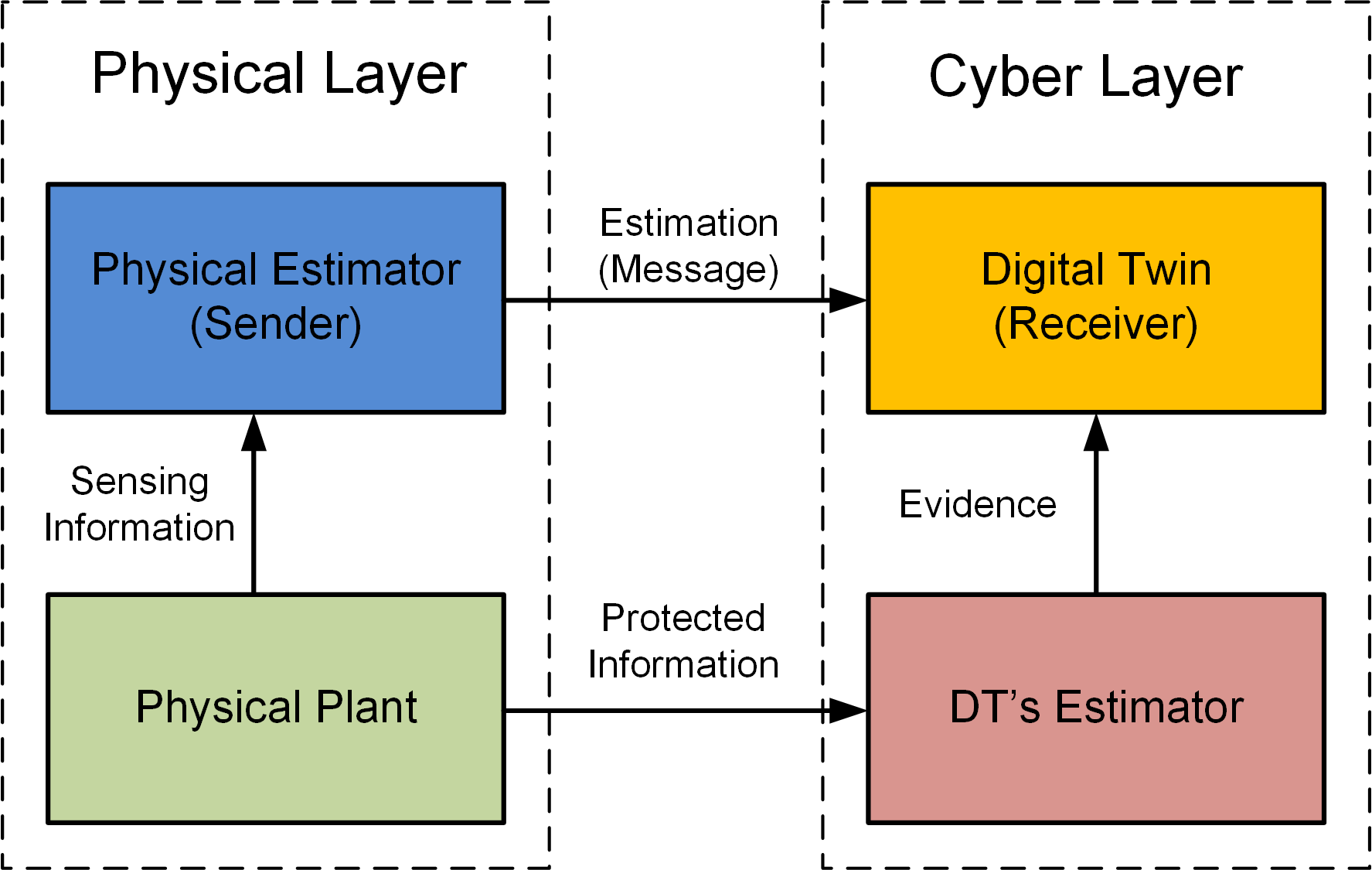}
  \caption{The Architecture of the Proposed SGE for a CPS: the physical estimator sends the estimation to the DT, which uses its secure evidence to verify the identity of the estimator.}\label{fig:sigGame}
  \vspace{-2mm}
\end{figure}

In an SGE, we have two players: one is the sender, and the other one is the receiver. The sender has a private identity, denoted by $\theta\in\Theta:=\{\theta_0, \theta_1\}$, where $\theta_0$ means the sender is benign, and $\theta_1$ means the sender is malicious. According to its identity, the sender will choose a message $m\in\calM$ and send it to the receiver. After observing the message, the receiver can choose an action $a\in\calA$. Action $a=1$ means that the receiver accepts the message, while $a=0$ means the receiver rejects the message. The sender and receiver have their own utility functions $U_i:\Theta\times\calM\times\calA\rightarrow\R$, for $i\in\{s, r\}$. Fig. \ref{fig:sigGame} illustrates how the data and information transmit in the proposed cyber-physical SGE.

In this paper, given a message $m\in\calM$, we assume that both players are aware of the corresponding detection results, i.e., $q=f_q(m)$. Hence, both players' can select the optimal strategies based on detection result $q$. To this end, we let $\sigma_s(f_q(m)|\theta)\in\Gamma_s$ and $\sigma_r(a|f_q(m))\in\Gamma_r$ be the mixed strategies of the sender and receiver, respectively. The spaces $\Gamma_s$ and $\Gamma_r$ are defined by
\begin{eqnarray}
  \Gamma_s&:=&
  \biggl\{ \sigma_s \biggl | \sum_{m\in\calM}\sigma_s(f_q(m)|\theta) = 1,
  \forall m, \ \sigma_s(f_q(m)|\theta) \ge 0 \biggl\}, \nonumber \\
  \Gamma_r&:=&
  \biggl\{ \sigma_r \biggl | \sum_{a\in\calA} \sigma_r(a| f_q(m)) = 1, \
  \forall a, \ \sigma_r(a|f_q(m)) \ge 0 \biggl\}. \nonumber
\end{eqnarray}
Note that formation of strategy $\sigma_s(f_q(m)|\theta)$ does not mean the sender can choose detection results directly. Instead, the sender can only choose message $m\in\calM$, which leads to a detection result $q$ based on function $f_q$, given by (\ref{eq:CSdetector}).

Given mixed strategies $\sigma_s$ and $\sigma_r$, we define players' expected utility functions as
\begin{eqnarray}
  \bU_s(\theta, \sigma_s, a) &:=& \sum_{q\in\calQ} \sigma_s(f_q(m)|\theta)
  U_s(\theta, m, a), \nonumber \\
  \bU_r(\theta, m, \sigma_r) &:=& \sum_{a\in\calA} \sigma_r(a|f_q(m))
  U_r(\theta, m, a). \nonumber
\end{eqnarray}
To find the optimal mixed strategies of both players, we identify a Perfect Bayesian Nash Equilibrium (PBNE) of the SGE in the following definition.
\begin{definition}\label{def:PBNE}
A PBNE of a SGE is a strategy profile ($\sigma^*_s, \sigma_r^*$) and a posterior belief $\pi(\theta)$ such that
\begin{eqnarray}
  \sigma_s(f_q(m)|\theta)&\in&\argmax_{\sigma_s\in\Gamma_s} \
  \sum_{a\in\calA} \sigma_r(a|f_q(m)) \bU_s(\theta, \sigma_s, a), \nonumber \\
  \sigma_r(a|f_q(m))&\in&\argmax_{\sigma_r\in\Gamma_r} \
  \sum_{\theta\in\Theta} \pi(\theta) \bU_r(\theta, m, \sigma_r), \nonumber
\end{eqnarray}
and the receiver updates the posterior belief using the Bayes' rule, i.e.,
\begin{eqnarray}
  \pi(\theta) &=& f_b(\pi'(\theta), f_q(m)) \nonumber \\
  &:=& \frac{\sigma_s(f_q(m)|\theta')\pi'(\theta)}
  {\sum_{\theta'\in\Theta}\sigma_s(f_q(m)|\theta')\pi'(\theta')}.
  \label{eq:belUp}
\end{eqnarray}
where $f_b:(0,1)\times \calQ \rightarrow (0,1)$ is the belief-update function, and $\pi'(\theta)$ is a prior belief of $\theta$.
\end{definition}

\begin{remark}
  Definition \ref{def:PBNE} identifies the optimal mixed strategies of the sender and receiver. One important thing is that at any PBNE, the belief $\pi(\theta)$ should be consistent with the optimal strategies, i.e., at the PBNE, belief $\pi(\theta)$ is independent of time $k$. Instead,  $\pi(\theta)$ should only depend on detection results $q\in\calQ$. Besides, we implement Bays's rule to deduce belief-update function (\ref{eq:belUp}).
\end{remark}

In a SGE, there are different types of PBNE. We present three types of PBNE in the following definition.
\begin{definition}
  (Types of PBNE)
  An SGE, defined by Definition \ref{def:PBNE}, has three types of PBNE:
  \begin{enumerate}
    \item Pooling PBNE: The senders with different identities use identical strategies. Hence, the receiver cannot distinguish the identities of the sender based on the available evidence and message, i.e., the receiver will use the same strategies with the different senders.
    \item Separated PBNE: Different senders will use different strategies based on their identities, and the receiver can distinguish the senders and use different strategies for different senders.
    \item Partially-Separated PBNE: different senders will choose different, but not completely opposite strategies, i.e.,
        \begin{eqnarray}
          \sigma_s(f_q(m)|\theta_0)
          &\neq& 1- \sigma_s(f_q(m)|\theta_1). \nonumber
        \end{eqnarray}
  \end{enumerate}
\end{definition}

\begin{remark}
  In the separated PBNE, the receiver can obtain the identity of the senders by observing a finite sequence of evidence and messages. However, in the other two PBNE, the receiver may not be able to distinguish the senders' identity.
\end{remark}

Note that in real applications, the CPS will run the SGE repeatedly, and generate a sequence of detection results $\calH_k:=\{q_0, q_1, \dots, q_{k} \}$. At time $k$, we define the posterior belief as
\begin{eqnarray}
  \pi_{k}(\theta) := \Pr(\theta|\calH_{k-1}).  \nonumber
\end{eqnarray}
Whenever there is a new detection result $q_k$, we can update the belief using $\pi_{k+1}(\theta) = f_b(\pi_k(\theta), q_k)$,
where function $f_b$ is defined by (\ref{eq:belUp}). Belief $\pi_{k+1}(\theta)$ will become a prior belief at time $k+1$.

To this end, we will use the SGE framework to capture the interactions between the physical layer and the DT. We will find the optimal defense strategy of the DT by finding the PBNE. In the next section, we define the utility functions explicitly and find the PBNE of the proposed SGE. Given the PBNE, we can identify the optimal defense strategies.



\section{Equilibrium Results of the Cyber SGE} \label{sec:main}

In this section, we aim to find the optimal defense strategy against a stealthy sensor attack. To this end, we first define the utility functions, which capture the profit earned by the players. Secondly, we identify the best response of the players when they observe or anticipate the other player's strategy. Finally, we present a PBNE under the players' best response and obtain an optimal defense strategy for the DT. We analyze the stability of the system under the stealthy attack.

\subsection{SGE Setup for the CPSs}

In this work, we use an SGE to capture the interactions between the physical estimator and the DT. In our scenario, the message set is just the estimation set, i.e., $\calM:=\R^{n_x}$. The DT monitors the estimation $m_k$ and chooses an action $a\in\calA:=\{0,1\}$. Action $a=1$ means the estimation passes the verification, while action $a=0$ means the verification fails, and the DT will send an alarm to the operators.

In the next step, we define the utility functions of both players, explicitly. Firstly, we define sender's utility functions $U_s(\theta, m, a)$. Since the sender has two identifies, we need to define two types of utility functions for the sender. The utility function with $\theta=0$ is defined by
\begin{eqnarray}
  U_s(\theta_0, m_k, a_k) := -\E[(m_k - x_k)^T(m_k - x_k)],
  \label{eq:uSender0}
\end{eqnarray}
In (\ref{eq:uSender0}), we can see that $U_s(\theta_0, m_k, a_k)$ is independent of action $a_k$, and maximizing $U_s(\theta_0, m_k, a_k)$ is equivalent to the estimation problem (\ref{eq:estProb}). Hence, given (\ref{eq:uSender0}), the benign sender always sends the true estimation result $\hx_k$, defined by (\ref{eq:Kalman}), regardless of action $a_k$.

For the malicious estimator, we define its utility function as
\begin{eqnarray}
  U_s(\theta_1, m_k, a_k) := \E[(m_k - x_k)^T(m_k - x_k) ]\cdot \bone_{\{a_k=1\}},
  \label{eq:uSender1}
\end{eqnarray}
where $\bone_{\{s\}}=1$ if statement $s$ is true. In (\ref{eq:uSender1}), we see that the motivation of the attacker is to deviate the system state as much as possible while remaining undiscovered.
However, the attacker's utility will be zero if the DT detects the attack.

Secondly, we define the DT's utility function. Note that the DT's utility function should depend on the identity of the sender. When the estimator is benign, i.e., $\theta=0$, the DT should choose $a_k=1$ to accept the estimation. When the estimator is malicious, i.e., $(\theta=1)$, the DT should choose $a_k=0$ to reject the estimation and send an alarm to the operators. Given the motivations, we define $U_r(\theta, \hx, a)$
\begin{eqnarray}
  U_r(\theta_0, m_k, a_k) &:=&
  - (\tx_k - m_k)^T\tQ_0(\tx_k - m_k)\cdot\bone_{\{a_k=0\}}, \nonumber \\
  U_r(\theta_1, m_k, a_k) &:=&
  - (\tx_k - m_k)^T\tQ_1(\tx_k - m_k)\cdot\bone_{\{a_k=1\}} . \nonumber
\end{eqnarray}
where $\tQ_0, \tQ_1\in\R^{n_x\times n_x}$ are positive-definite matrices. The weighting matrices will affect the receiver's defense strategy. A large value of $\tr(\tQ_1)$ will lead to a conservative strategy, while a large value of $\tr(\tQ_0)$ will lead a radical one. Readers can receive more details in Proposition \ref{thm:bestDT}.

In the next subsection, we analyze the behaviors of the players and obtain the best-response strategies. Note that function $U_r(\theta, m_k, a_k)$ is deterministic. The reason is that the DT can observe $\hx_k$ and $\tx_k$ at time $k$, explicitly. However, the physical estimator cannot observe $x_k$ at time $k$.

\subsection{Best Response of the Players and a PBNE of the SGE}

We first analyze the best response of the DT. Given belief $\pi_{k}(\theta)$, message $m_k$, and detection result $q_k$, we present the following theorem to identify DT's best response.
\begin{proposition}\label{thm:bestDT}
(DT's Best Response) Given $q_k=f_q(m_k)$, the DT will choose $a_k$ according to the following policy,
\begin{eqnarray}
  \sigma^*_r(a_k=1|q_k) &=&
  \left\{
    \begin{array}{ll}
      1, & \text{~if~} q_k\ne 2,
      \pi_k(\theta_0) \ge \beta; \vspace{1mm}\\
      0, & \text{~if~} q_k\ne 2,
      \pi_k(\theta_0) < \beta;   \vspace{1mm}\\
      0, & \text{~if~} q_k = 2; \\
    \end{array}
  \right. \label{eq:bestResDT} \\
  \sigma^*_r(a_k=0|q_k) &=& 1 - \sigma^*_r(a_k=1|q_k),
\end{eqnarray}
where $\beta$ is defined by
\begin{eqnarray}
  \beta:= \frac{(\tx_k - m_k)^T\tQ_1(\tx_k - m_k)}
  {(\tx_k - m_k)^T(\tQ_0 + \tQ_1)(\tx_k - m_k)}. \label{eq:defBeta}
\end{eqnarray}

\end{proposition}
\begin{IEEEproof}
  Note that
  \begin{eqnarray}
     \E[U_r(\theta, m_k, a_k=1)|q_k]  &\ge&
       \E[U_r(\theta, m_k, a_k=0)|q_k] \nonumber \\
     \Leftrightarrow  \quad
    a_k=1 \text{~if~} \pi_k(\theta_0) &\ge& \beta, \nonumber
  \end{eqnarray}
  where $\beta$ is defined by (\ref{eq:defBeta}). This completes the proof.
\end{IEEEproof}

\begin{remark}
  Given Proposition \ref{thm:bestDT}, we note that the DT uses a pure strategy since it can make its decision after observing detection result $q_k$ and message $m_k$.
\end{remark}

In the next step, we consider the best response of the estimator. If the estimator is benign, i.e., $\theta=\theta_0$, the optimal estimation should be (\ref{eq:Kalman}). Therefore, the optimal utility of the benign estimator is given by
\begin{eqnarray}
  U_s(\theta_0, \hx_k, a_k) &=&
  \E[(\hx_k - x_k)^T(\hx_k - x_k)]
  = \tr(\hP_k), \nonumber
\end{eqnarray}
where $\tr(P)$ is the trace of matrix $P$. The following theorem shows the optimal mixed strategy of the benign estimator.
\begin{proposition}\label{thm:bestBS}
  (Best Response of the Benign Estimator)
  Given the DT's best response (\ref{eq:bestResDT}), the optimal mixed strategy of the benign estimator, i.e., $\theta=\theta_0$, is given by
  \begin{eqnarray}
    \sigma^*_s(f_q(m_k)=0| \theta_0) &=& F_{\chi}(\rho_1, n_x), \label{eq:bestBS0} \\
    \sigma^*_s(f_q(m_k)=1 | \theta_0) &=& F_{\chi}(\rho_2, n_x) - F_{\chi}(\rho_1, n_x), \label{eq:bestBS1} \\
    \sigma^*_s(f_q(m_k)=2| \theta_0) &=& 1 - F_{\chi}(\rho_2, n_x), \label{eq:bestBS2}
  \end{eqnarray}
  where $\hx_k$ is defined by (\ref{eq:Kalman}), $F_{\chi}(\rho, n):\R_{+}\rightarrow[0,1]$ is the Cumulative Distribution Function (CDF) of the Chi-square distribution with $n\in\Z_{+}$ degrees.
\end{proposition}

\begin{IEEEproof}
  Note that the benign estimator will choose $m_k = \hx_k$, defined by (\ref{eq:Kalman}). According to definition (\ref{eq:ChiSqu}), we know that $(\tx_k - \hx_k)$ follows a Chi-square distribution with $n_x$ degrees. Hence, we have
  \begin{eqnarray}
    && \Pr(\chi^2_k\le \rho_1) = F_{\chi}(\rho_1, n_x), \nonumber \\
    && \Pr(\chi^2_k > \rho_2]) =
    1- F_{\chi}(\rho_2, n_x), \nonumber \\
    && \Pr(\chi^2_k\in (\rho_1, \rho_2]) =
    F_{\chi}(\rho_2, n_x) -  F_{\chi}(\rho_1, n_x). \nonumber
  \end{eqnarray}
  Combining the above equations with Chi-square detector (\ref{eq:CSdetector}) yields mixed strategies (\ref{eq:bestBS0})-(\ref{eq:bestBS2}).
\end{IEEEproof}

\begin{remark}
Note that the benign estimator always choose the optimal estimation \ref{eq:Kalman}. However, from DT's perspective in this game, the real mixed strategies of the benign estimator are (\ref{eq:bestBS0})-(\ref{eq:bestBS2}) because of uncertainty introduced by the noises.
\end{remark}

From the perspective of the malicious estimator, it needs to select $\sigma_s(f_q(m_k)|\theta_1)$ such that $\pi_{k}(\theta_0) \ge \beta$. Given the attackers' incentive, we obtain the following theorem.
\begin{proposition}\label{thm:bestAT}
  (Best Response of the Malicious Estimator)
  \begin{eqnarray}
    && \sigma^*_s(f_q(\xi_{k,1})=0| \theta_1) =
       F_{\chi}(\rho_1, n_x), \label{eq:bestAT0} \\
    && \sigma^*_s(f_q(\xi_{k,2})=1 | \theta_1) =
    1- F_{\chi}(\rho_1, n_x), \label{eq:bestAT1}
  \end{eqnarray}
  where $\xi_{k, 1}, \xi_{k,2}$ are the solutions to the following problems:
  \begin{eqnarray}
    \xi_{k,1}\in\argmax_{m\in\calM_{\rho_1}(\tx_k)} && U_s(\theta_1, m, a_k=1),
    \label{eq:maxUs1} \\
    \xi_{k,2}\in\argmax_{m\in\calM_{\rho_2}(\tx_k)} && U_s(\theta_1, m, a_k=1),
    \label{eq:maxUs2}
  \end{eqnarray}
  with spaces $\calM_{\rho_1}(\tx_k)$ and $\calM_{\rho_2}(\tx_k)$ defined by
  \begin{eqnarray}
    \calM_{\rho_1}(\tx_k) &:=& \biggl\{ m\in\R^{n_x} \biggl|
    \|\tx_k - m\|^2_{\Sigma^{-1}_{\phi}}
    \le \rho_1 \biggl\}, \nonumber  \\
    \calM_{\rho_2}(\tx_k) &:=& \biggl\{ m\in\R^{n_x} \biggl|
    \|\tx_k - m\|^2_{\Sigma^{-1}_{\phi}}
     \in(\rho_1, \rho_2] \biggl\}. \nonumber
  \end{eqnarray}
\end{proposition}

\begin{IEEEproof}
  Firstly, the attacker has no incentive to choose $m_k\notin\calM_{\rho_1}(\tx_k) \cup \calM_{\rho_2}(\tx_k)$ because its utility will be zero. Secondly, the attacker aims to choose the mixed strategy $\sigma^*_s(q_k=1 | \theta_1)$ as large as possible since it can make a higher damage to the system. Then, we show that the optimal mixed strategy of the attacker is given by (\ref{eq:bestAT0}) and (\ref{eq:bestAT1}).
  To do this, based on (\ref{eq:belUp}), we consider the following belief update:
  \begin{eqnarray}
     && \pi_{k+1}(\theta_0) \nonumber \\
    &&= \frac{\sigma^*_s(q_k\neq0| \theta_0) \pi_{k}(\theta_0)}
    {\sigma^*_s(q_k\neq0| \theta_0) \pi_{k}(\theta_0)
    + \sigma_s(q_k\neq0| \theta_1) (1-\pi_{k}(\theta_0))} \nonumber \\
    &&= \frac{\sigma^*_s(q_k\neq0| \theta_0) \pi_{k}(\theta_0)}
    {\Delta\sigma_s(q_k\neq0)  \pi_{k}(\theta_0) +  \sigma_s(q_k\neq1| \theta_1)},
   \label{eq:lambdaU}
  \end{eqnarray}
  where $\Delta\sigma_s(q_k\neq 0)$ is defined by
  \begin{eqnarray}
  \Delta\sigma_s(q_k\neq0):= \sigma^*_s(q_k\neq0| \theta_0)
                 - \sigma_s(q_k\neq0| \theta_1). \nonumber
  \end{eqnarray}

  Rearranging (\ref{eq:lambdaU}) yields that
  \begin{eqnarray}
    \frac{\pi_{k+1}(\theta_0)}{\pi_{k}(\theta_0)}
  = \frac{\sigma^*_s(q_k\neq0| \theta_0) \pi_{k}(\theta_0)}
    { \Delta\sigma_s(q_k\neq0) \pi_{k}(\theta_0) \
    +  \sigma_s(q_k\neq0| \theta_1)}. \nonumber
  \end{eqnarray}
  Given that $\pi_{k}(\theta_0)\in(0,1]$, we have
  \begin{eqnarray}
  \left\{
    \begin{array}{cc}
      \frac{\pi_{k+1}(\theta_0)}{\pi_{k}(\theta_0)}>1,
      & \text{~if~} \Delta\sigma_s(q_k\neq0) > 0; \vspace{1mm}\\
      \frac{\pi_{k+1}(\theta_0)}{\pi_{k}(\theta_0)}=1,
      & \text{~if~} \Delta\sigma_s(q_k\neq0) = 0; \vspace{1mm}\\
      \frac{\pi_{k+1}(\theta_0)}{\pi_{k}(\theta_0)}<1,
      & \text{~if~} \Delta\sigma_s(q_k\neq0) < 0. \vspace{1mm}\\
    \end{array}
  \right. \nonumber
  \end{eqnarray}
  When $\pi_{k}(\theta_0)\in[\beta, 1)$, the attacker has to choose $\Delta\sigma_s(q_k\neq0) = 0$ to maintain the belief at a constant. Otherwise, the belief will decrease continuously. When the belief $\pi_k(\theta_0)$ stays lower than $\beta$, the DT will send an alert to the operators. Hence, the optimal mixed strategies of the malicious estimator are given by (\ref{eq:bestAT0}) and (\ref{eq:bestAT1}).
\end{IEEEproof}

Given the results of Propositions \ref{thm:bestDT}, \ref{thm:bestBS}, and \ref{thm:bestAT}, we present the following theorem to characterize a unique pooling PBNE.

\begin{theorem}\label{thm:poolPBNE}
  (The PBNE of the Proposed SGE)
  The proposed cyber SGE has a unique pooling PBNE. At the PBNE, the optimal mixed strategies of the benign and malicious sender are presented by (\ref{eq:bestBS0})-(\ref{eq:bestBS2}) and (\ref{eq:bestAT0})-(\ref{eq:bestAT1}). The DT has a pure strategy defined by (\ref{eq:bestResDT}). At the PBNE, belief $\pi^*_k(\theta_0)\in[\beta, 1)$ is a fixed point of function $f_b$, i.e.,
  \begin{eqnarray}
    \pi^*_k(\theta_0) = f_b(\pi^*_k(\theta_0), q_k), \text{~for~} q_k\in\calQ.
    \nonumber
  \end{eqnarray}
\end{theorem}

\begin{IEEEproof}
  We first show the existence of the pooling PBNE. Suppose that both estimators use strategies (\ref{eq:bestBS0})-(\ref{eq:bestBS2}), (\ref{eq:bestAT0})-(\ref{eq:bestAT1}), respectively, and the DT uses (\ref{eq:bestResDT}). Then, no player has incentive to move since these are already the optimal strategies. Besides, for any $\theta\in\Theta, q_k\in\calQ$, we note that
  \begin{eqnarray}
    \pi_{k+1}(\theta) &=& f_b(\pi^*_k(\theta), q_k) = \pi^*_{k}(\theta), \nonumber
  \end{eqnarray}
  where $f_b$ is defined by (\ref{eq:belUp}). Hence, $\pi^*_k(\theta)$ is a fixed point of function $f_b$, and the belief remain at $\pi^*_k(\theta)$, which means the belief stays consistently with the optimal strategies of the sender and receiver. Hence, the proposed strategies pair $(\sigma^*_s, \sigma^*_r)$ is a PBNE.

  Secondly, we show that pooling PBNE is unique. We note that the DT and benign estimator have no incentive to move since they already choose their best strategies. In Proposition \ref{thm:bestAT}, we already show that the attacker cannot change its mixed strategies. Otherwise, the belief cannot remain constant. Hence, pooling PBNE is unique.
\end{IEEEproof}

\begin{remark}
  Theorem \ref{thm:poolPBNE} shows that the SGE admits a unique pooling PBNE, which means that an intelligent attacker can use its stealthy strategies to avoid being detected by the DT.
\end{remark}

In the next subsection, we will analyze the stability of the system under the stealthy attack. Besides, we will also evaluate the loss caused by the attack.

\subsection{Estimated Loss Under the Stealthy Attack}

In the previous subsection, we have shown the PBNE in which the attacker can use a stealthy strategy to pass the verification of the DT. In this subsection, we will quantify the loss under the attack. Before presenting the results, we need the following lemma.
\begin{lemma}
Given $\rho_i$ and $\xi_{k,i}$, for $i\in\{1,2\}$, we have the following relationship:
\begin{eqnarray}
  \rho_i\lambda_{\max}(\Sigma_{\phi})
  &\ge&  (\tx_k - \xi_{k,i})^T(\tx_k - \xi_{k,i}), \text{~for~} i=1,2, \nonumber
\end{eqnarray}
where $\lambda_{\max}(\Sigma)$ is the greatest eigenvalue of matrix $\Sigma$.
\end{lemma}
\begin{IEEEproof}
  Firstly, we note that $U_s$ is strictly convex in $m_k$. The solution to problem (\ref{eq:maxUs1}) and (\ref{eq:maxUs2}) must stay at the boundary.
  Hence, we have
\begin{eqnarray}
  \rho_i &=& (\tx_k - \xi_{k, i})^T\Sigma^{-1}_{\phi}(\tx_k - \xi_{k,i}) \nonumber \\
  &\ge& \frac{(\tx_k - \xi_{k,i})^T(\tx_k - \xi_{k,i})}
  {\lambda_{\max}(\Sigma_{\phi})} \label{eq:rhoSigma}
\end{eqnarray}
Rearranging (\ref{eq:rhoSigma}) yields that
\begin{eqnarray}
  \rho_i\lambda_{\max}(\Sigma_{\phi})
  &\ge&  (\tx_k - \xi_{k,i})^T(\tx_k - \xi_{k,i}). \nonumber
\end{eqnarray}
This completes the proof.
\end{IEEEproof}

Considering different estimators, we define two physical cost functions $J_0$ and $J_1$, i.e.,
\begin{eqnarray}
  J_0 &:=& \lim_{N\rightarrow \infty}\E\biggl\{ \frac{1}{N}
  \sum_{k=0}^{N-1}\biggl[ x^T_kQx_k + \mu^T(m_k)R\mu(m_k)\biggl]
  \biggl|\theta_0 \biggl\}, \nonumber \\
  J_1 &:=& \lim_{N\rightarrow \infty}\E\biggl\{ \frac{1}{N}
  \sum_{k=0}^{N-1}\biggl[ x^T_kQx_k + \mu^T(m_k)R\mu(m_k)\biggl]
  \biggl|\theta_1\biggl\}. \nonumber
\end{eqnarray}
We define a loss function $\Delta J := J_1 - J_0$ to quantify the loss caused by the stealthy sensor attack. Given pooling PBNE defined by Theorem \ref{thm:poolPBNE}, we provide an upper-bound of $\Delta J$ in the following theorem.

\begin{theorem}\label{thm:bJ}
  (Bounded Loss)
  The proposed framework can guarantee stability of the CPSs, and the value of function $\Delta J$ is bounded by a constant, i.e.,
  \begin{eqnarray}
  \Delta J &=& J_1 - J_0 \nonumber \\
  &\le&  \alpha_1 \tr(\tP^*) - \alpha_0\tr(\hP^*)
  + \alpha_1\rho_1\lambda_{\max}(\Sigma_{\phi})F_{\chi}(\rho_1, n_x)
  \nonumber \\
  &+& \alpha_1\rho_2\lambda_{\max}(\Sigma_{\phi})(1-F_{\chi}(\rho_1, n_x)),
  \label{eq:jbound}
\end{eqnarray}
where $\alpha_0$ and $\alpha_1$ are defined by
\begin{eqnarray}
   \alpha_0 &:=& \frac{\lambda_{\min}(R_K)(\lambda_{\min}(G) - \lambda_{\min}(R_K))}{\lambda_{\min}(G)}, \nonumber \\
    \alpha_1 &:=& \frac{\lambda_{\max}(R_K)(\lambda_{\max}(R_K)
  + 2\lambda_{\max}(G))}{\lambda_{\max}(G)},  \nonumber
\end{eqnarray}
and $R_K:= K^TRK, G:=Q+R_K$; $\lambda_{\max}(W)$ and $\lambda_{\min}(W)$ are the greatest and smallest eigenvalues of matrix $W$.
\end{theorem}

\begin{IEEEproof}
Firstly, we note that
\begin{eqnarray}
  &&\E\biggl[\|x_k\|^2_Q  + \|K\hx_k\|^2_{R}\biggl]  \nonumber \\
  &=& \E\biggl[\|x_k\|^2_{G} + 2x^T_kR_K\he_k +  \|\he_k\|^2_{R_K}\biggl] \nonumber \\
  &\ge& \E\biggl[\biggl\|\sqrt{\lambda_{\min}(G) } x_k + \frac{\lambda_{\min}(R_K)}{\sqrt{\lambda_{\min}(G)}}\he_k \biggl\|^2
   \nonumber \\
  &+& \frac{\lambda_{\min}(R_K)(\lambda_{\min}(G) - \lambda_{\min}(R_K))}{\lambda_{\min}(G)} \|\he_k\|^2\biggl]
  \ge \alpha_0 \tr(\hP_k), \nonumber
\end{eqnarray}
Using the above inequality, we observe that
\begin{eqnarray}
   J_0
  &=& \lim_{N\rightarrow\infty}
   \frac{1}{N}\sum_{k=0}^{N-1} \E
  \biggl[ \|x_k\|^2_{Q} + \|K\hx_k\|^2_{R}\biggl] \nonumber \\
  &\ge& \lim_{N\rightarrow\infty}
   \frac{1}{N}\sum_{k=0}^{N-1} \alpha_0\tr(\hP_k)
    = \alpha_0\tr(\hP^*), \label{eq:J0}
\end{eqnarray}
where $\hP^*$ is defined by (\ref{eq:KalCov}). Secondly, we also note that
\begin{eqnarray}
  &&  \E\biggl[ x^T_kQx_k + \mu^T(\xi_{k,i})R\mu(\xi_{k,i})\biggl]\nonumber\\
 &=& \E\biggl[ \|x_k\|^2_{Q} + \|x_k + \te_k
  + \xi_{k,i} - \tx_k \|^2_{R_K} \biggl] \nonumber \\
  &\le& \E\biggl [\|x_k\|^2_{G} +
  \biggl(2x^T_kR_K\te_k + 2x^T_kR_K(\xi_{k,i} - \tx_k) \biggl)  \nonumber \\
  &+&  2\te^T_kR_K(\xi_{k,i} - \tx_k)
  + \|\xi_{k,i} - \tx_k\|^2_{R_K} + \|\te_k\|^2_{R_K}\biggl] \nonumber \\
  &\le& \E \biggl[3\|x_k\|^2_{G} +
  \frac{\lambda^2_{\max}(R_K)}{\lambda_{\max}(G)}\biggl( \|\te_k\|^2+
  \|\xi_{k,i} - \tx_k\|^2\biggl) \nonumber \\
  &+& 2\lambda_{\max}(R_K)\|\te_k\|^2
   + 2\lambda_{\max}(R_k)\|\xi_{k,i} - \tx_k\|^2\biggl] \nonumber \\
  &\le& 3\|x_k\|^2_{G} + \alpha_1 \tr(\tP_k)
  + \alpha_1\rho_i\lambda_{\max}(\Sigma_{\phi}), \label{eq:inEq1}
\end{eqnarray}
We complete the squares to deduce the second inequality of (\ref{eq:inEq1})
Similarly, we have
\begin{eqnarray}
  J_1
  &=& \lim_{N\rightarrow\infty}
 \frac{1}{N}\sum_{k=0}^{N-1}\biggl\{
  F_{\chi}(\rho_1, n_x)\E\biggl[ \|x_k\|^2_{Q}
  + \|\xi_{k, 1} \|^2_{R_K}\biggl] \nonumber \\
  &+& \biggl (1 - F_{\chi}(\rho_1, n_x)\biggl)
  \E\biggl[ \|x_k\|^2_{Q} + \|\xi_{k, 1} \|^2_{R_K}\biggl] \biggl\} \nonumber \\
  &\le& \underbrace{\lim_{N\rightarrow\infty} \frac{3}{N}\sum_{k=0}^{N-1} \|x_k\|^2_{G}}_{=0}
  + \alpha_1\rho_1\lambda_{\max}(\Sigma_{\phi})F_{\chi}(\rho_1, n_x) \nonumber \\
  &+& \alpha_1\rho_2\lambda_{\max}(\Sigma_{\phi})(1-F_{\chi}(\rho_1, n_x))
  + \alpha_1 \tr(\tP^*), \label{eq:J1}
\end{eqnarray}
where $\tP^*$ is defined by (\ref{eq:DTcov}). Combining inequalities (\ref{eq:J0}) and (\ref{eq:J1}) yields inequality (\ref{eq:jbound}). Hence, the system is stable, and the impact of the attack is bounded by a constant.
\end{IEEEproof}

\begin{remark}
  Theorem \ref{thm:bJ} shows that the difference between $J_0$ and $J_1$ is bounded, i.e., the stealthy estimation attack cannot deviate the system to an arbitrary point even if the attacker has an infinite amount of time.
\end{remark}

In the next subsection, we will use an application to evaluate the performance of the proposed defense strategies.

\section{Simulation Results}\label{sec:simRes}

In this section, we use a two-link Robotic Manipulator (RM) to investigate the impact of the estimation attacks. In the experiments, we use different case studies to analyze the performance of the proposed defense framework.

\begin{figure}[thp]
  \centering
  \includegraphics[width=3.2cm]{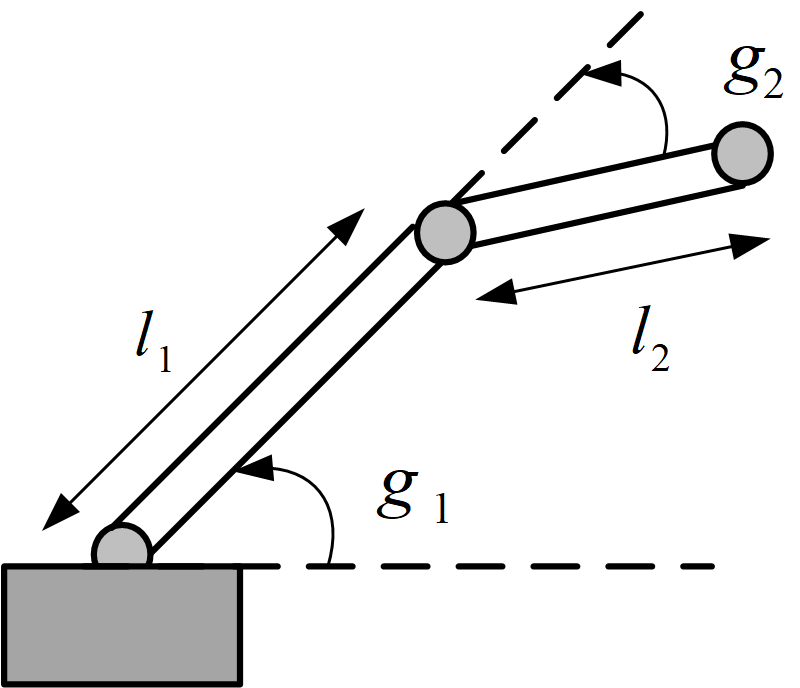}
  \caption{The Dynamic Model of a Two-Link Robotic Manipulator (RM): the RM has two links and moves in a two-dimensional space. }\label{fig:twoLinkModel}
  \vspace{-5mm}
\end{figure}

\begin{figure*}
\begin{center}
\begin{minipage}[b]{0.31\linewidth}
  \centering
  \includegraphics[scale=0.4]{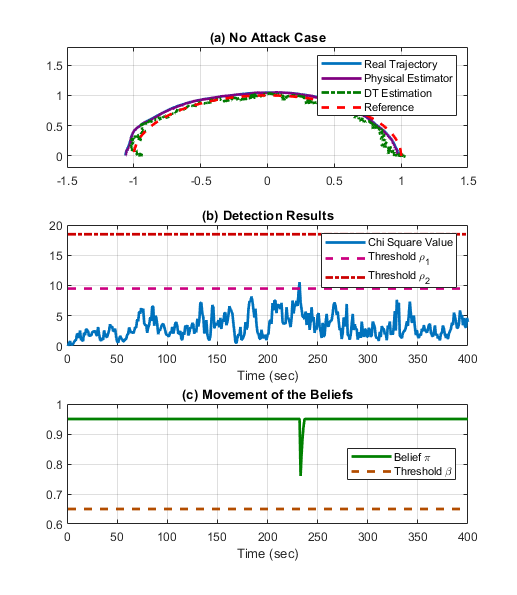}
  \vspace{-2mm}
  \caption{No-Attack Case: (a) the system trajectory, physical estimation and DT's estimation; (b) the Chi-square value; (c) DT's belief $\pi(\theta_0)$. }\label{fig:simRes1}
\end{minipage}
\hspace{0.1cm}
\begin{minipage}[b]{0.31\linewidth}
  \centering
  \includegraphics[scale=0.4]{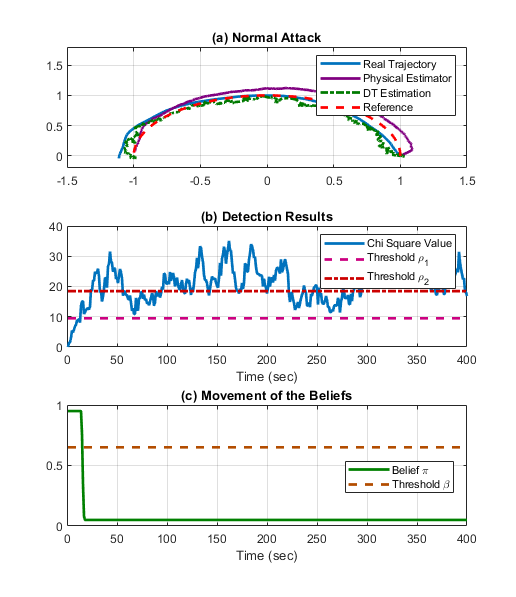}
  \vspace{-2mm}
  \caption{Normal-Attack Case: (a) the system trajectory, physical estimation and DT's estimation; (b) the Chi-square value; (c) DT's belief $\pi(\theta_0)$.
   }\label{fig:simRes2}
\end{minipage}
\hspace{0.1cm}
\begin{minipage}[b]{0.31\linewidth}
  \centering
  \includegraphics[scale=0.4]{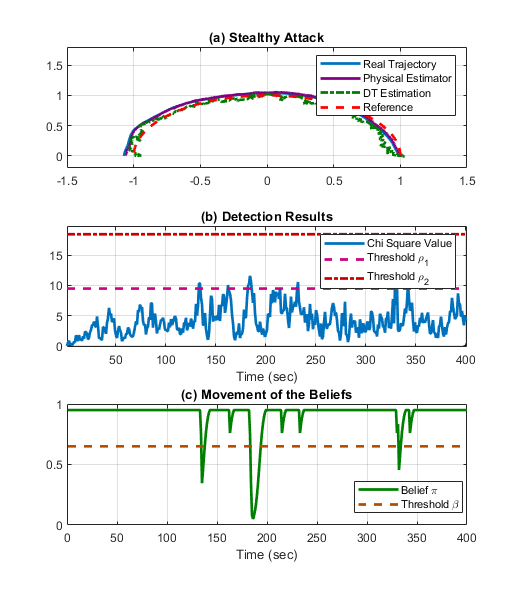}
  \vspace{-2mm}
  \caption{Stealthy-Attack Case: (a) the system trajectory, physical estimation and DT's estimation; (b) the Chi-square value; (c) DT's belief $\pi(\theta_0)$.
  }\label{fig:simRes3}
\end{minipage}
\end{center}
\vspace{-5mm}
\end{figure*}

\subsection{Experimental Setup}

Fig. \ref{fig:twoLinkModel} illustrates the physical structure of the two-link RM. Variables $g_1$ and $g_2$ are the angular positions of Links 1 and 2. We summarize the parameters of the RM in Table II.
\begin{table}[thp]
\caption{Parameters of the Robotic Manipulator}
\centering
\begin{tabular}{|c|c|c|}
\hline
Parameter & Description                 & Value           \\ \hline
$l_1$     & Length of Link 1            & 0.6 m            \\ \hline
$l_2$     & Length of Link 2            & 0.4 m            \\ \hline
$r_1$     & Half Length of Link 1       & 0.3 m            \\ \hline
$r_2$     & Half Length of Link 2       & 0.2 m            \\ \hline
$\eta_1$     & Mass of Link 1              & 6.0 kg           \\ \hline
$\eta_2$     & Mass of Link 2              & 4.0 kg           \\ \hline
$I_1$     & Inertia of Link 1 on z-axis & 1 kg$\cdot$m$^2$ \\ \hline
$I_2$     & Inertia of Link 2 on z-axis & 1 kg$\cdot$m$^2$ \\ \hline
\end{tabular}
\vspace{-2mm}
\end{table}

Let $g=[g_1, g_2]^T$ be the angular vector and $\tau = [\tau_1, \tau_2]^T$ be the torque input. According to the Euler-Lagrange Equation, we obtain the dynamics of the two-link RM as
\begin{eqnarray}
  M(g)\ddg + S(g, \dg)\dg = \tau, \label{eq:twoLink}
\end{eqnarray}
where matrices $M(g)$ and $S(g, \dg)$ are defined by
\begin{eqnarray}
  M(g) &:=&
  \left[
    \begin{array}{cc}
      a + b\cos(g_2) & \delta + b\cos(g_2) \\
      \delta + \cos(g_2) & \delta \\
    \end{array}
  \right], \nonumber \\
  S(g, \dg) &:=&
  \left[
    \begin{array}{cc}
      -b\sin(g_2)\dg_2 & -b\sin(g_2)(\dg_1+\dg_2) \\
      b\sin(g_2)\dg_1 & 0 \\
    \end{array}
  \right], \nonumber\\
  a &:=& I_1 + I_2 + \eta_1 r_1^2 + \eta_2(l^2_1 + r^2_2),  \nonumber \\
  b &:=& \eta_2l_1r_2, \qquad \delta := I_2 + \eta_2r^2_2.  \nonumber
\end{eqnarray}
To control the two-link RM, we let $\tau$ be $\tau := M(g)a_g + S(g, \dg)\dg$, where $a_q\in\R^2$ is the acceleration that we need to design. Note that $M(g)$ is positive-definite, i.e., $M(g)$ is invertible. Hence, substituting $\tau$ into (\ref{eq:twoLink}) yields that
\begin{eqnarray}
  M(g)\ddg = M(g)a_g \ \Rightarrow \ \ddg = a_g. \nonumber
\end{eqnarray}

Let $p\in\R^{2}$ be the position of RM's end-effector. We have
\begin{eqnarray}
  \ddot{p} = H(g)\ddg + \dot{H}(g)\dg = H(g)a_g+ \dot{H}(g)\dg, \label{eq:pState}
\end{eqnarray}
where $H(g)$ is the Jacobian matrix. Then, we substitute $a_g:= H^{-1}(g)(u - \dot{H}(g))$ into (\ref{eq:pState}), arriving at $\ddot{p} = u$. Let $x = [p^T, \dot{p}^T]^T$ be the continuous-time state. Then, we obtain a continuous-time linear system $\dx = A_cx + B_cu$. Given a sampling time $\Delta T>0$, we discretize the continuous-time system to obtain system model (\ref{eq:sysModel}). We let $y_k$ and $z_k$ be
\begin{eqnarray}
  y_k = x_k + v_k, \quad z_k = p(k\Delta T) + d_k
\end{eqnarray}
We assume that the DT uses security-protected cameras to identify the position of the end-effector.

In the experiments, we let the RM to draw a half circle on a two-dimensional space. The critical parameters are given by
\begin{eqnarray}
  && \beta = 0.65, \ n_x = 4, \ \rho_1 = 9.49, \ \rho_2 = 18.47, \nonumber \\
  && F_{\chi}(\rho_1, n_x) = 0.95, \quad F_{\chi}(\rho_2, n_x) = 0.999. \nonumber
\end{eqnarray}
We have three case studies: a no-attack case, a normal-attack case, and a stealthy-attack case. In the normal-attack case, the attacker is not aware of the defense strategies and deviate the system from the trajectory, directly. In the last case, the attacker aims to tamper the estimation without being detected.

Figures \ref{fig:simRes1}, \ref{fig:simRes2}, and \ref{fig:simRes3} illustrate the simulation results of the case studies. In Fig. \ref{fig:simRes1} (a), we can see that the RM can track the trajectory smoothly when there is no attack. However, we note that DT's estimation is worse than the physical estimation, which coincides with our expectation. Figures \ref{fig:simRes1} (b) and (c) show the value of the Chi-square and the belief of the DT. In the no-attack case, the Chi-square detector will remain silent with a low false alarm rate, and the belief stays at a high level.In Fig. \ref{fig:simRes2} (a), the attackers deviate the system without considering the detection. Even though DT's estimation is not accurate, the attacker cannot tamper that. Therefore, the detector will rapidly locate the attack and send alarms to the operators. The belief of $\theta_0$ will remain at the bottom line. In Fig. \ref{fig:simRes3}, differently, the stealthy attackers know the defense strategies and try to maintain the Chi-square value below threshold $\rho_1$. However, the behavior mitigates the impact of the attack, which also coincides with the result of Theorem \ref{thm:bJ}.

\begin{figure}[thp]
  \centering
  \includegraphics[width=6.8cm]{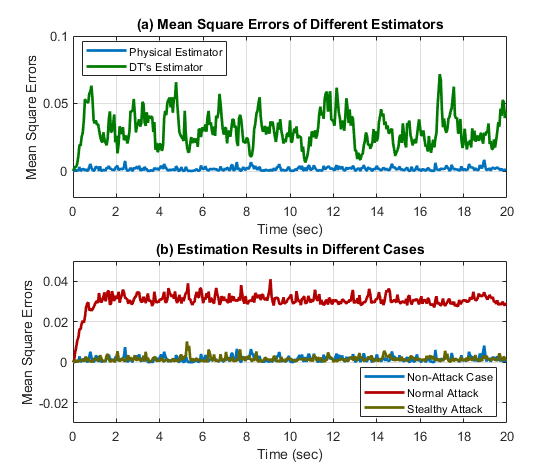}
  \caption{The Comparison of the Mean Square Error (MSE): (a) the comparison between the MSE of physical estimation and DT's estimation; (b) the MSE of different case studies.}\label{fig:simRes4}
  \vspace{-4mm}
\end{figure}

Figure \ref{fig:simRes4} illustrates the Mean Square Errors (MSE) of different cases. Figure \ref{fig:simRes4} (a) presents that the MSE of the physical estimator is much smaller than the DT's estimator, i.e., the physical estimator can provide more accurate sensing information. However, in Figure \ref{fig:simRes4} (b), we can see that the attacker can deviate the physical estimation to a significant MSE. Besides, under the DT's supervision, the stealthy attacker fails to generate a large MSE. The above results show that the proposed defense mechanism succeeds in mitigating the stealthy attacker's impact.

\section{Conclusions} \label{sec:con}

In this paper, we have considered a stealthy estimation attack, where an attack can modify the estimation results to deviate the system without being detected. To mitigate the impact of the attack on physical performance, we have developed a Chi-square detector, running in a Digital Twin (DT). The Chi-square detector can collect DT's observations and the physical estimation to verify the identity of the estimator. We have used a Signaling Game with Evidence (SGE) to study the optimal attack and defense strategies. Our analytical results have shown that the proposed framework can constrain the attackers' ability and guarantee the stability.

\vspace{-1mm}
\bibliographystyle{IEEEtran}
\bibliography{IEEEabrv,ICCPS_bib}

\end{document}